\documentclass[amsmath,amssymb,amsbsy,reprint,prb,preprintnumbers,showpacs,superscriptaddress]{revtex4-2}
\usepackage{graphicx,color}
\usepackage{dcolumn}
\usepackage{bm}
\usepackage{braket}
\usepackage{mathtools}
\usepackage{ulem}
\usepackage[breaklinks,colorlinks=true,linkcolor=blue,urlcolor=blue,citecolor=blue]{hyperref}
\usepackage{times}
\usepackage{physics}
\usepackage{latexsym}
\usepackage{amsmath, amssymb}
\usepackage{mathtools}
\usepackage{multirow}
\usepackage{hyperref}

\newcommand{\be}{\begin{eqnarray}}
\newcommand{\ee}{\end{eqnarray}}

\renewcommand{\theequation}{\arabic{equation}}

\begin{document}

\title{
Mixed-state phase structure of gauge-Higgs subsystem codes under logical-preserving decoherence
}
\date{\today}
\author{Yoshihito Kuno} 
\affiliation{Graduate School of Engineering Science, Akita University, Akita 010-8502, Japan}

\author{Ikuo Ichinose} 
\thanks{A professor emeritus}
\affiliation{Department of Applied Physics, Nagoya Institute of Technology, Nagoya, 466-8555, Japan}


\begin{abstract}
Some of lattice-gauge-theory models, in particular gauge-Higgs model (GHM), can be regarded and work as a subsystem code. 
This work studies the effect of local-gauge-symmetric decoherence on the GHM from the perspective of the subsystem code. 
We clarify the global phase diagram of the subsystem code. 
In particular, the decoherence induces an unconventional critical mixed state, where the logical information is preserved 
but the rest of the system exhibits mixed state criticality. 
For a fixed point, the decohered subsystem code is understood by the “gauging out” prescription. 
By mapping the GHM to the toric code subject to decoherence, we can understand the properties of the subsystem code.
We further discuss and investigate the robustness of the logical space of the subsystem code. 
Although this kind of subsystem code can be produced by using any bulk mixed state in the GHM, its robustness is a subtle problem due to 
the mixed critical gauge qubits. 
We consider some specific unitary for examining the robustness of the stored quantum information.
For dynamical unitary perturbations described by interactions between the logical qubit and gauge qubits, 
the deformation of the subsystem code drastically depends on the initial mixed state of the gauge qubits. 
\end{abstract}


\maketitle
\section{Introduction}
Lattice gauge theory~\cite{Fradkin1979,Kogut1979} was invented to describe the matter-gauge interactions
in elementary-particle physics, in particular,
strong interactions \cite{Wilson1974} and quark confinement mechanism.
It is also a non-perturbative formalism for understanding 
spontaneous breaking of global ($0$-form) symmetry as well as higher-form symmetries in the modern perspective.
Among various lattice models, the lattice gauge-Higgs model (LGHM) is versatile.
That is, besides the high-energy physics context, the LGHM plays an important role since it provides a deep understanding of various 
condensed matter phenomena~\cite{Fradkin_text,IchinoseMatsui2014} such as (high-temperature) superconductivity~\cite{Lee1992} 
and deconfined quantum critical point~\cite{Senthil2004}.
Recently, some studies \cite{Borla2021,Verresen2024} claimed that Higgs phase in the LGHM can be regarded as a symmetry-protected topological (SPT) phase \cite{Pollmann2012}, 
and by introducing spatial boundaries, i.e., considering a cylinder system, the Higgs and confinement regimes are distinguishable by boundary operators 
characterizing nature of the edge modes \cite{Verresen2024}, contrary to the common belief.

The LGHM has also been recognized as an important model in quantum information, that is, some limit of $Z_2$-gauge version of the LGHM 
is nothing but the toric code (TC) with explicit local-gauge symmetries~\cite{Kitaev2003,Wang2003,Pachos2012}. 
The quantum phase appeared in the TC works as a quantum memory \cite{Dennis2001,Kitaev2003,Arakawa2004,Ohno2004}. 
The phase possesses a topological order \cite{Wen2007}, which is an on-going attractive state of matter in condensed matter physics. 
Then, from the quantum information perspective, the TC is a topological stabilizer code \cite{Gottesman1998,Nielsen2011} 
and its quantum memory can be fault-tolerant \cite {Kitaev2003}.
Interestingly, many properties of the TC as quantum memory can be deeply understood from the $Z_2$ LGHM model \cite{Dennis2001,Kitaev2003}. 

From the perspective of quantum information science, the LGHM is of further significance as it can be regarded as another type of quantum memory
named subsystem code~\cite{Poulin2005,knill2006,Bacon2006}. 
Interestingly enough, a recent study~\cite{Wildeboer2022} indicated that generic LGHMs with open boundaries 
become a subsystem code if one sets some suitable open boundary conditions. 
In the open boundary system, degenerate eigenstates of the LGHM behave as encoded qubits with gauge qubits 
since one finds boundary logical operators that are an anticommutative pair \cite{Wildeboer2022}, leading to a quantum memory. 
[Here, we comment that ``gauge qubit", ``gauge operator" and ``gauge group" have nothing to do with gauge symmetry/invariance of gauge theory.
See later explanation.]
Each term of the model Hamiltonian is an element of the gauge operator \cite{Ellison2023,KI2023}, and the gauge invariant operators 
in the gauge theory are generators of a stabilizer group, which is the center of the gauge group~\cite{Poulin2005,Bacon2006,Wildeboer2022,KI2023}. 
The subsystem code may offer stronger protection against environmental disturbances than a stabilizer code due to the redundancy provided 
by so-called gauge qubits.
Moreover, the Hilbert-space structure of subsystem code may give an efficient error correction route as discussed in Ref.~\cite{Poulin2005}: 
The syndrome measurements in the subsystem code often involve products of fewer measured qubits than 
those in stabilizer code, which can make the error-correction process easier to implement~\cite{Bacon2006,Bravyi2013}.

In a realistic situation, the effect of environment is inevitable. 
In the literature, the effects of environment on stabilizer code, including the TC, have been extensively studied. 
In particular, how decoherence described by a quantum channel \cite{Nielsen2011,lidar2020} affects the TC  
\cite{JYLee2023,Bao2023,Fan2024,Sang2025,Zhang2024strong} has been studied in quantum information and condensed matter communities. 
How the logical space in the TC changes under a change in bulk topological order has been investigated \cite{Wang2025,Sohal2025}. 
However, the situation is much less clear for the subsystem code. 
For example,
(i)how decoherence affects the LGHM from the perspective of the subsystem code, 
(ii)how the subsystem codes can coexist with nontrivial mixed states,
(iii)how logical (topological) stability of the subsystem code changes under the variation of the gauge-qubit state;
These remain to be clarified through a detailed investigation on a concrete model.

In this work, we study the effect of decoherence on an extended LGHM, elucidating the structure between the gauge-qubit state 
and the encoded logical space under a gauge-operator decoherence. 
We discuss how the subsystem code coexists with various mixed phases and what global mixed-state phase structure is in the LGHM 
under the decoherence.
By mapping the LGHM to the TC and its related statistical mechanical model recently proposed in \cite{JYLee2023},
we draw the global mixed-state phase diagram, indicating that the mixed state with the subsystem-code structure is rich, 
that is, the decoherence induces various mixed phases with the subsystem logical qubit intact.
We find that at a fixed point, the decohered subsystem code changes through the “gauging out” prescription~\cite{Ellison2023,Sohal2025}. 
We also emphasize that the subsystem code with critical mixed states exists. 
This mixed state is an interesting mixed state from the viewpoint of many-body phases. 

Furthermore, by introducing specific dynamical perturbations that couple the logical and gauge-qubit degrees of freedom, 
we examine the stability of the subsystem code and show how it is degraded differently depending on the properties of the gauge-qubit sector. 
In particular, for the gauge-qubit state near criticality, the subsystem code is fragile. 
The result indicates that subsystem codes coexisting with a mixed state in the vicinity of critical points are less stable 
against environmental disturbance compared to those with stable gauge-qubit states 
(corresponding to``gapped'' mixed states in the sense of Markov gap proposed in \cite{Sang2025}).

The rest of this paper is organized as follows. 
In Sec.~II, we explain the LGHM, the specific boundary conditions employed in this work, and its symmetry aspects. A mapping to the TC is introduced.  
Then, we discuss how the LGHM becomes a subsystem code in detail (If the reader wants to know the essence of subsystem code in detail, see Refs.~\cite{Poulin2005,Wildeboer2022}). In Sec.~III, we introduce a type of decoherence and discuss the effect of the decoherence, and describe the global mixed-phase structure with the subsystem code. 
There, the global mixed-phase structure is described by employing the mapping to the TC\cite{Wildeboer2022}, the gauging out procedure \cite{Ellison2023,Sohal2025}, and statistical mechanical picture \cite{JYLee2023}.
In Sec.~IV, we numerically demonstrate the effect of the decoherence on the subsystem code. In Sec.~V, we study the stability of the decohered subsystem code to a perturbation. We elucidate that the mixed state of the gauge qubits gives a significant impact on the stability of the subsystem code.  
Section VI is devoted to discussion and conclusion.

\section{Model}
We first introduce the target lattice model, which is a lattice-gauge-theory model \cite{Kogut1979}
composed of two kinds of $Z_2$ gauge fields and coupled two kinds of $Z_2$ matter fields. 
Then, we show how it is connected to the TC with open boundaries. 
We further explain the aspects of subsystem code \cite{Poulin2005,Wildeboer2022} by explicitly observing the structure of the Hilbert space.  

\begin{figure}[t]
\begin{center} 
\includegraphics[width=8cm]{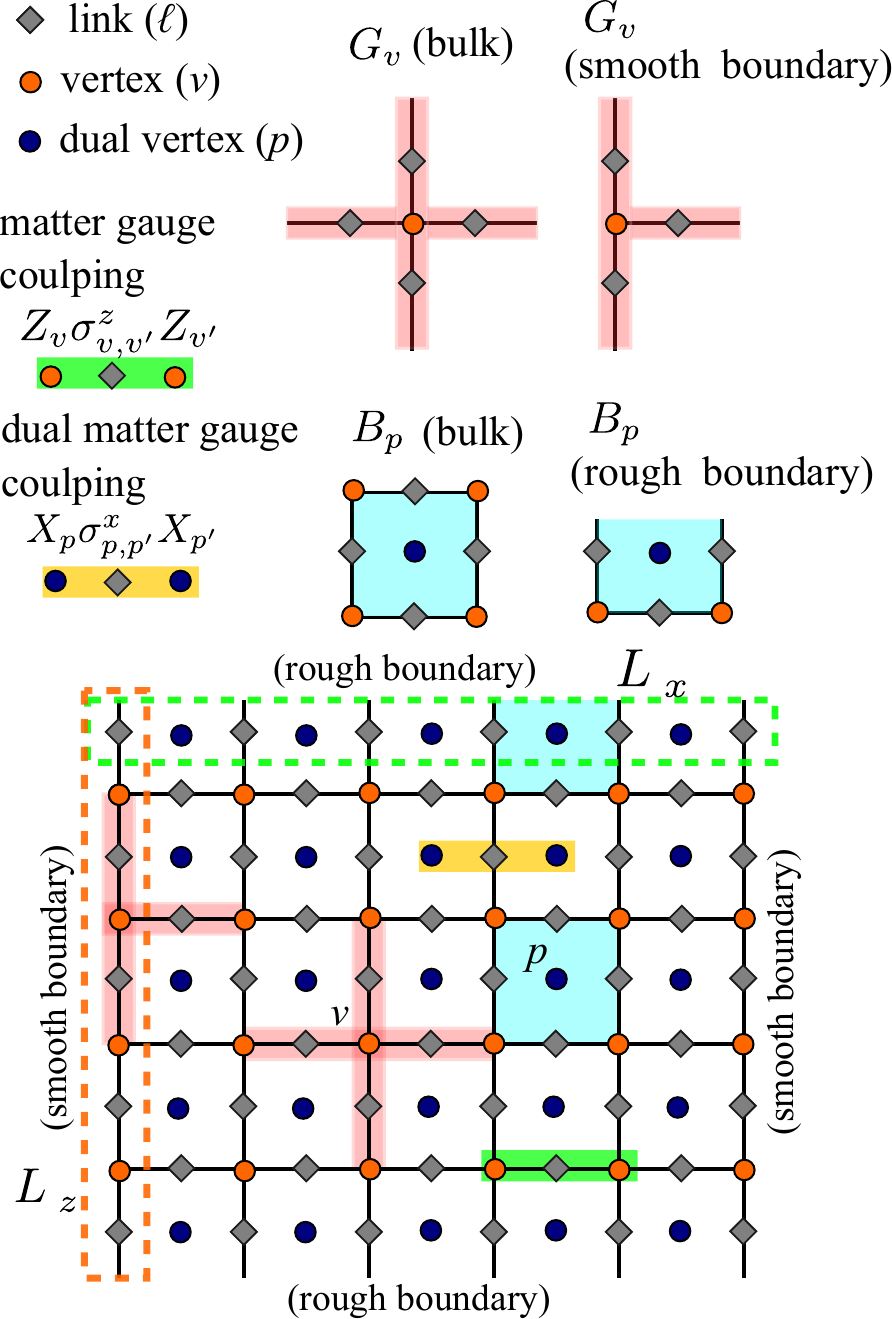}  
\end{center} 
\caption{Schematic figures of the lattice of the LGHM with rough and smooth boundaries. 
The green dashed box represents the top rough boundary, where the bare logical operator $L_x$ is defined and the orange dashed box 
represents the left smooth boundary, where the bare logical operator $L_z$ is defined.}
\label{Fig1}
\end{figure}

\subsection{Full-dual description of $Z_2$ lattice gauge-Higgs model}
In this work, we focus on the full-dual description of LGHM, which explicitly satisfies 
the electric-magnetic duality including mater fields and is also a maximally-gauged system of the TC. 
The Hamiltonian is given as
\begin{eqnarray}
H_{\rm LGHM}&=&-\sum_{v}X_v 
-\sum_{(p,p')}JX_{p}\sigma^x_{p,p'}X_{p'}\nonumber\\
&-&\sum_{p}Z_p-\sum_{(v,v')}JZ_{v}\sigma^z_{v,v'}Z_{v'}.
\label{LGHM}
\end{eqnarray}
where $J$ is a parameter of the matter-gauge coupling, and for the physical subspace of the Hilbert space (called original physical Hilbert space), 
the following double gauge-invariant conditions [Gauss laws] are imposed,
\begin{eqnarray}
G_v|\psi\rangle=|\psi\rangle,\:\:
B_p|\psi\rangle=|\psi\rangle,
\label{const}
\end{eqnarray}
where
\begin{eqnarray}
&&G_v=X_v\prod_{\ell_{v} \in v} \sigma^x_{\ell_{v}}\equiv X_v\tilde{G}_v, \nonumber  \\
&&B_p=Z_p\prod_{\ell_{p} \in p} \sigma^z_{\ell_{p}}\equiv Z_p\tilde{B}_p,
\label{GvBp}
\end{eqnarray}
and $\ell_{v} \in v$ stands for links emanating from vertex (site) $v$, and 
$\ell_{p} \in p$ for links composing plaquette (box) $p$. 
The schematic image for the lattice structure and each terms 
of the Hamiltonian $H_{\rm LGHM}$ are shown in Fig.~\ref{Fig1}.
The $Z_2$-electric matter is defined on each vertex $v$, $(X_v,Z_v)$, and its magnetic dual, $(X_p,Z_p)$, on each dual vertex $p$ 
(i.e., plaquette of the original lattice), where $X_v (Z_v)$ stands for 
the Pauli matrix $\sigma^x_v (\sigma^z_v)$, and similarly for $X_p (Z_p)$.
On the other hand, the $Z_2$ gauge field is defined on links and denoted by $(\sigma^x_\ell, \sigma^z_\ell)$, 
$\sigma^z_{v,v'}$ denote a gauge field on link connecting neighboring vertices $v$ and $v'$, and $\sigma^x_{p,p'}$ 
denote a gauge variable on link connecting neighboring dual vertices $p$ and $p'$. 
The gauge field $\sigma^x_\ell$ is related to the electric field $\hat{E}_\ell$ as $\sigma^x_\ell=e^{i\pi \hat{E}_\ell}$, 
and $\sigma^z_\ell$ to the vector potential $\hat{A}_\ell$ as $\sigma^z_\ell=e^{i\pi \hat{A}_\ell}$, 
and eigenvalues are $\{ 0, 1 \}$ for both the operators~\cite{Kogut1975,Kogut1979}.

In later discussion, we show that the system $H_{\rm LGHM}$ in Eq.~(\ref{LGHM}) is nothing but the genuine $Z_2$
gauge-Higgs model and also the perturbed TC with modifications suitable for the present issue,
both of which play important roles in various research fields.
We emphasize that contrary to the ordinary $Z_2$ LGHM \cite{Kogut1979,Fradkin1979},
$H_{\rm LGHM}$ contains ``flux-matter" field $(X_p, Z_p)$, and this makes $H_{\rm LGHM}$ have the exact full-duality such as 
$\{\sigma^z,\sigma^x,X_v,Z_v,Z_p,X_p\} \rightarrow \{\sigma^x,\sigma^z,Z_p,X_p,X_v,Z_v\}$ [plus diagonal half translations]. 
This full-duality is maintained for the system with rough and smooth boundaries introduced later on.

In this work, we consider a square lattice with specific boundaries named rough and smooth boundaries. 
For both boundaries, the forms of $G_v$ and $B_p$ are explicitly shown (See Fig.~\ref{Fig1}), 
both of which are composed of three links with a vertex and a plaquette, respectively. 
This open LGHM has a specific symmetry property. 
The model with the above open boundary conditions enjoys four symmetries, 
generators of which are given by,
\begin{eqnarray}
\hat{P}&=&\prod_{v}X_v,\:\:
\hat{S}_Z=\prod_{p}Z_p,\\
W_{\gamma}&=&\prod_{\ell \in \gamma}\sigma^z_{\ell},\:\:
H_{\gamma}=\prod_{\ell \in \gamma}\sigma^x_{\ell}.
\label{symmetry}
\end{eqnarray}
$\hat{P}$ is the parity of the total $Z_2$ electric charge, corresponding to the global spin flip on each vertex,
and similarly, $\hat{S}_Z$ is the parity of the total $Z_2$ magnetic flux per plaquette.
The boundary hopping terms are forbidden by the $\hat{P}$ and $\hat{S}_Z$ symmetries. 
These satisfies $[H_{LGHM},\hat{P}]=[H_{LGHM},\hat{S}_Z]=0$.  
$\hat{P}$ and $\hat{S}_Z$ can be regarded as global topological symmetries \cite{Wildeboer2022}, 
whereas $W_{\gamma}$ and $H_{\gamma}$ are one-form symmetry, which has been extensively studied recently \cite{Gaiotto2015, McGreevy2023}. 
There, $\gamma$ denotes a contractible closed-loop. (In the lattice gauge theory, these operators are Wilson and ’t Hooft loops 
operators \cite{Wilson1974,THOOFT19781}.)
When $\gamma$ is open, the corresponding string operators create the electric or magnetic excitations at the endpoints 
and hence are not symmetry generators.
Although the two matter-gauge coupling terms ($J$-terms) in the Hamiltonian $H_{\rm GHM}$ [Eq.~(\ref{HGHM})] explicitly break 
the one-form symmetries, it was shown that the higher-form symmetry is generally robust and 
give a non-trivial effect on the dynamics of the system \cite{Gaiotto2015,McGreevy2023}. 

\subsection{Connection to toric code with rough and smooth open boundaries}
The LGHM can be mapped onto the TC including local perturbations ($J$-terms) \cite{Kitaev2003}. 
Here, this mapping enables us to well-capture the physical properties of $H_{\rm LGHM}$. In particular, the effect of a decoherence can be well-captured and smoothly understood by following the TC under a decoherence. 

We here consider the following unitary transformation \cite{Wildeboer2022}
\begin{eqnarray}
U_v&=&H\biggl(\prod_{v}\prod_{\ell\in v} (CZ)_{v,\ell}\biggr)H,\\
U_p&=&H\biggl(\prod_{p}\prod_{\ell\in p} (CZ)_{p,\ell}\biggr)H,
\end{eqnarray}
where $H$ is the Hadamard transformation on each link and 
$(CZ)_{i,j}$ is a controlled $Z$-gate for the site $i$ and link $j$.
Applying the above transformation for all $v$ and $p$ as $U_vU_pH_{\rm LGHM}(U_vU_p)^\dagger$, 
the LGHM is mapped to the following effective disentangled model, 
\begin{eqnarray}
H_{\rm LGHM}&&\xrightarrow[\text{plaquette fixing}]{U_{p}} H_{\rm GH},\nonumber\\
&&\xrightarrow[\text{gauge fixing}]{U_{v}U_{p}} H_{\rm TC},\nonumber\\
H_{\rm GH}&&=-\sum_v X_v-\sum_{\ell \notin \mbox{smooth}} J \sigma^x_{\ell }\nonumber \\
&-&\sum_p \tilde{B}_p-\sum_{(v,v')}JZ_v\sigma^z_{v,v'}Z_{v'},  \nonumber  \\  
H_{\rm TC}&&=-\sum_{v}\tilde{G}_v  -\sum_{\ell \notin \mbox{rough}}J\sigma^z_{\ell}\nonumber\\
&-&\sum_{p}\tilde{B}_p-\sum_{\ell \notin \mbox{smooth}}J\sigma^x_{\ell}.
\label{HTC}
\end{eqnarray}
Here, we have applied the unitary transformation to the LGHM ($Z_p=+1$) first, 
and then carried out a gauge fixing to the unitary gauge ($X_v=+1$). 
The $J$-terms originate from the matter-gauge couplings in $H_{\rm LGHM}$. 
Hamiltonian $H_{\rm GH}$ is the genuine $Z_2$ gauge Higgs model with one important modification, i.e., 
the electric term $\sum J\sigma^x_\ell$ does not contain the smooth boundary terms as dictated by the original 
flux-hopping terms in $H_{ \rm LGHM}$ [Eq.~(\ref{LGHM})].
On the other hand, the model $H_{\rm TC}$ is nothing but the perturbed TC \cite{Kitaev2003} 
including local perturbations ($J$-terms) and with rough and smooth boundaries.
$H_{\rm TC}$ with periodic boundary conditions and also $H_{\rm TC}$ having the full $J$-terms under the standard open boundary conditions 
have been studied in previous works from the viewpoint of quantum information \cite{Wildeboer2022,KI2023_sc}.

We further see how the symmetry operators $\hat{P}$ and $\hat{S}_Z$ are also transformed to find,
\begin{eqnarray}
\hat{P}\xrightarrow{U_{v}U_{p}}\tilde{P}=\prod_{\ell\in \mbox{rough}} \sigma^x_{\ell},\:\:\:
\hat{S}_Z\xrightarrow{U_{v}U_{p}}\tilde{S}_Z=\prod_{\ell\in \mbox{ smooth}} \sigma^z_{\ell}.\nonumber
\end{eqnarray}
These forms of the symmetry operators lead to ``uniqueness'' for logical operators in subsystem code \cite{Wildeboer2022,KI2023_sc}. 


\subsection{Subsystem code}
In this subsection, we briefly explain the subsystem code and its relation to the stabilizer code. 
This can be understood by observing the Hilbert space structure of the LGHM. We start with showing the Hilbert space structure of the LGHM in detail. 
As for Ref.~\cite{Wildeboer2022}, it is proposed that the LGHM becomes a subsystem code if one sets suitable open boundary conditions. 
Each term of the Hamiltonian (\ref{LGHM}) is an element of the gauge group \cite{Ellison2023} in the subsystem-code literature \cite{Poulin2005} 
and the present gauge group is given by
$$
\mathcal{G}=\langle\{X_{p}\sigma^x_{p,p'}X_{p'}\},\{Z_{v}\sigma^z_{v,v'}Z_{v'}\},\{X_v\},\{Z_p\} \rangle,
$$
where we omit $\hat{P}$, $\hat{S}_Z$ operator since these operators are always fixed in this work. 
Here, the stabilizer group can be obtained by taking the center of the gauge group $\mathcal{G}$
$$
\mathcal{Z}(\mathcal{G})=\langle \{G_v\},\{B_p\}\rangle.
$$
This stabilizer set $\mathcal{Z}(\mathcal{G})$ makes the space decomposed. The original Hilbert space $\mathcal{H}_{phys}$ becomes
\begin{eqnarray}
\mathcal{H}_{phys}&=&\mathcal{C}\oplus \bar{C},\nonumber \\
G_v|\psi\rangle&=&|\psi\rangle,\:\:
B_p|\psi\rangle=|\psi\rangle,\:\:\mbox{if}\; |\psi\rangle \in C.\nonumber
\end{eqnarray}
The space $\mathcal{C}$ is the code space and the space $\bar{C}$ is orthogonal to $\mathcal{C}$. 
Throughout this work, we focus on the space $\mathcal{C}$.

Furthermore, the LGHM has topological symmetries, $\hat{P}$ and $\hat{S}_Z$. These symmetries induce $Z_2$ charge sectors, raveled by four different sectors as $(P,S_Z)=(\pm 1,\pm 1)$.  
The space $\mathcal{C}$ is further decomposed as
\begin{eqnarray}
\mathcal{C}=\bigoplus_{P,S_Z=\pm 1}C^{(P,S_Z)}.
\end{eqnarray}
Each space $C^{(P,S_Z)}$ is further factorized \cite{Poulin2005} as
$$
C^{(P,S_Z)}=\mathcal{L}\otimes \bar{\mathcal{L}}^{(P,S_Z)},
$$
where $\mathcal{L}$ is a logical subsystem independent to the sector label $(P,S_Z)$ and $\bar{\mathcal{L}}^{(P,S_Z)}$ is 
a gauge-junk space depending on the sector label $(P,S_Z)$. 
In the subsystem code, a logical encoded qubit resides in $\mathcal{L}$ while the degree of freedom in $\bar{\mathcal{L}}^{(P,S_Z)}$ 
is called ``gauge qubits''. 
In this work, we consider the sector sum of $\bar{\mathcal{L}}^{(P,S_Z)}$, $\mathcal{L}_G\equiv\bigoplus_{P,S_Z=\pm 1}\bar{\mathcal{L}}^{(P,S_Z)}$ 
called ``gauge-junk space''.

One also recognizes the presence of one pair of bare logical operators for $\mathcal{L}$ as
\begin{eqnarray}
L_x=\prod_{\ell\in \mbox{top rough}} \sigma^x_{\ell},\:\:\:
L_z=\prod_{\ell\in \mbox{left smooth}} \sigma^z_{\ell}.
\label{logical_OP_1Q}
\end{eqnarray}
This satisfies $\{L_x,L_z\}=0$. 
Thus, a single qubit can be encoded in the system as a subsystem code. 
In particular, the ground state manifold of $H_{\rm LGHM}$ is two degenerate orthogonal pair states corresponding to the single logical qubit. 
In addition, we note that the bare logical operators $L_x$ and $L_z$ are invariant under the transformation $U_{v}U_{p}$ introduced 
in the previous section. 
In what follows, we focus on a fixed sector $(P,S_Z)$ within the space $\mathcal{C}$. 
Thus, we shall omit the label of the $Z_2$ charge symmetry sector $(P,S_Z)$.

Note that, although the above Hilbert space decomposition exists, the $Z_2$ charge symmetry does not induce degeneracy; 
that is, the states extracted from each $(P,S_Z)$ sector have different energies from the Hamiltonian of the LGHM. 
In contrast, the anticommutating logical operators $L_x$ and $L_z$ give the degeneracy structure. 
When we observe the pure ground state of the LGHM, in a broad regime of the model parameters, 
the ground state exhibits two-fold degeneracy respecting the logical space with a fixed symmetry sector $(P,S_Z)$ selected. 
For an explicit construction of the sectors, see \cite{KI2023}.

\begin{figure}[t]
\begin{center}
\includegraphics[width=6.5cm]{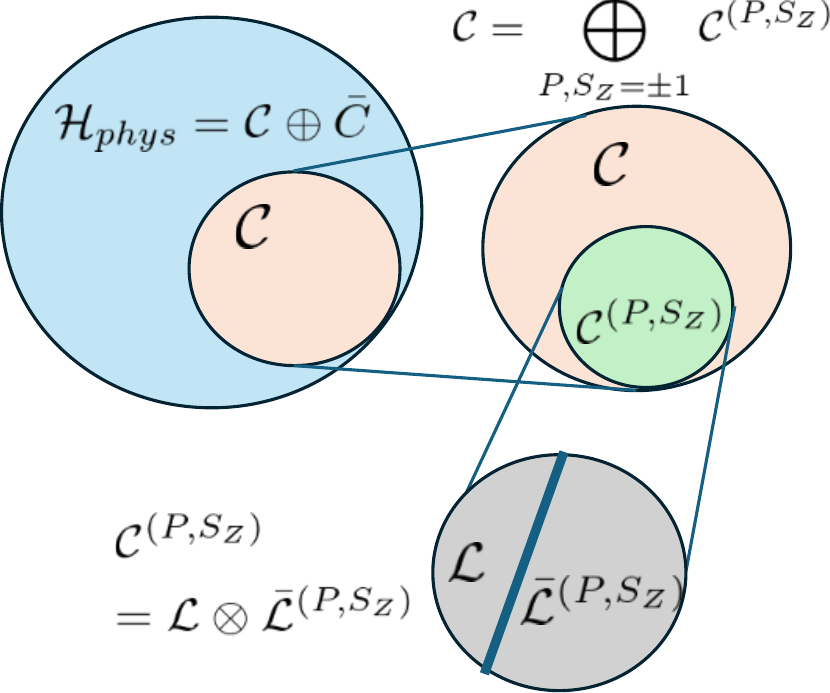}
\end{center}
\caption{Schematic image of the Hilbert space structure of the LGHM. 
The original Hilbert space $\mathcal{H}_{phys}$ is decomposed in the stabilizer group $\mathcal{Z}(\mathcal{G})$. 
The stabilized space $\mathcal{C}$ is also composed by $Z_2$ charge sector. 
Furthermore, each $Z_2$ charge sector in the stabilized space $\mathcal{C}$ is factorized with logical space $\mathcal{L}$ and gauge-junk space 
$\bar{\mathcal{L}}^{(P,S_Z)}$.
}
\label{Fig2}
\end{figure}

\subsection{Toric code perspective of subsystem code}
The LGHM is mapped into the TC with rough and smooth boundaries. 
The logical code in the LGHM is preserved; that is, under the mapping, the logical space is preserved. 
Through the mapping, the bare logical operators do not change, $U_vU_pL_{x(z)}(U_vU_p)^\dagger=L_{x(z)}$. 
Strictly, the logical code turns into a stabilizer code, not subsystem code, since each term is a ``stabilizer'', commutative with each other. 
This comes from the gauge fixing, $Z_v=+1$ and $X_p=+1$. 
However, the properties of the subsystem code in the LGHM are well-understood via the TC picture. 

Using the mapping from the LGHM to the TC, we comment on the relationship of the Hilbert space structure between them. 
Indeed for a fixed $Z_2$ charge sector, the structure transforms as 
\begin{eqnarray}
(\mathcal{L}\otimes \bar{\mathcal{L}}^{(P,S_Z)})\oplus \bar{C}^{(P,S_Z)}
\xrightarrow[\text{gauge fixing}]{U_{v}U_{p}} \mathcal{L}_{TC}\oplus \bar{C}_{TC}^{(P,S_Z)}.
\end{eqnarray}
The space $\bar{C}^{(P,S_Z)}$ orthogonal to the stabilized space $C$ in $\mathcal{H}_{\rm LGHM}$ is reduced by
the gauge fixing $Z_v=+1$ and $X_p=+1$ in the LGHM side. 
That is, the spaces $\mathcal{L}_{TC}$ and $\bar{C}_{TC}^{(P,S_Z)}$ correspond to $\mathcal{L}$ and $\bar{\mathcal{L}}^{(P,S_Z)}$, respectively. 
Note that by the gauge fixing the degree of gauge qubit $\bar{\mathcal{L}}^{(P,S_Z)}$ is frozen, but the original gauge degree of freedom comes to be in $\bar{C}^{(P,S_Z)}_{TC}$.

For the later discussion in the following section and Sec.V, we here introduce a gauge operator in the LGHM denoted by $\hat{O}_G$, 
which is generally constructed by picking up the operators in $\mathcal{G}$. 
From the Hilbert space structure of the LGHM, the operator $\hat{O}_G$ acts trivially on the logical subsystem 
$\mathcal{L}$, whereas non-trivially on the orthogonal complement $\bar{C}^{(P,S_Z)}$. 
Similarly, under the unitary mapping $U_vU_p$ [Eqs.~(6) and (7)], the operator $\hat{O}_G$ acts trivially on the logical subsystem 
$\mathcal{L}_{TC}$ and non-trivially on the orthogonal complement 
$\bar{C}^{(P,S_Z)}_{\mathrm{TC}}$ as
\begin{equation}
U_vU_p\, \hat{O}_G\, (U_vU_p)^\dagger 
= I_{\mathcal{L}_{TC}} \otimes \hat{O}_{\bar{C}^{(P,S_Z)}_{TC}}.
\end{equation}
Hence, the decoherence channel composed of such gauge operators 
affects exclusively the gauge-junk part of the Hilbert space, 
while leaving the logical code space invariant.

\section{Mixed state subsystem code}
We study the effect of a specific type of decoherence, and we choose gauge operators for decoherence. 
We discuss what the entire states of the LGHM are for various parameters and the strength of decoherence. 

\subsection{Gauge-operator decoherence channel}
This work focuses on a gauge-operator type of decoherence, described by the following quantum channel \cite{Nielsen2011}
\begin{eqnarray}
&&\mathcal{E}^{g}=\prod_{(p,p')}\mathcal{E}^g_{(p,p')},\nonumber\\
&&\mathcal{E}^g_{(p,p')}[\rho]=(1-p_x)\rho+p_x (X_{p}\sigma^x_{p,p'}X_{p'})\rho (X_{p}\sigma^x_{p,p'}X_{p'}),\nonumber\\
\label{HGHM}
\end{eqnarray}
where $p_x$ is the strength of decoherence taking $0\leq p_x\leq 1/2$ and the operator $X_{p}\sigma^x_{p,p'}X_{p'}$ is an element of $\mathcal{G}$. 
The gauge-invariant decoherence channel $\mathcal{E}^{g}$ gives dynamics to the subsystem code state with a fixed 
$(P,S_Z)$ sector, that is, the gauge qubits are affected by $\mathcal{E}^{g}$, resulting that the gauge qubits evolve into a mixed state. 
Then, the $(P,S_Z)$ sector itself remains unchanged. 

While the logical space $\mathcal{L}$ is preserved even in any strength of the decoherence $\mathcal{E}^{g}$, the state in the gauge group $\mathcal{G}$ is drastically affected. The initial pure state is set as $|\psi_L\rangle \otimes |\psi_{G}\rangle$ where $|\psi_L\rangle \in \mathcal{L}$ and $|\psi_L\rangle \in \bar{\mathcal{L}}^{(P,S_Z)}$, 
\begin{eqnarray}
\rho_0=(|\psi_L\rangle \langle \psi_L|) \otimes (|\psi_{G}\rangle \langle \psi_G|).
\end{eqnarray}
Then, formally, the initial state is decohered by $\mathcal{E}^{g}$ as
\begin{eqnarray}
\mathcal{E}^g[\rho_0]=(|\psi_L\rangle \langle \psi_L|) \otimes \mathcal{E}^g[|\psi_{G}\rangle \langle \psi_G|].
\end{eqnarray}
That is, the gauge qubit is mixed. 
The mixed state is rich with the subsystem logical space preserved. This becomes an interesting many-body mixed state.

\subsection{Decoherence in zero-Higgs coupling limit}
We show that for both $J=0$ and $p_x=1/2$ limits, the code can be understood in the framework of the ``gauging out'' procedure \cite{Ellison2023} 
proposed recently for mixed states from the viewpoint of subsystem code~\cite{Sohal2025}.

For $J=0$ and $p_x=0$ limits, 
the system is described by a stabilizer  
$$
\mathcal{S}_{J=0}=\langle\{X_v\},\{Z_p\} \rangle.
$$
The ground state of the LGHM is a simple product.
Under $J=0$ and $p_x=1/2$ limits, the gauge qubits are mixed and the gauge operator does not commute with the group $\mathcal{S}_{J=0}$.
Then, a mixed stable state can be deduced by gauging out \cite{Sohal2025}. 
As for the gauging out, we consider the gauge group constituted by combining the stabilizer group with the decohered operators as \cite{Sohal2025}
\begin{eqnarray}
\mathcal{G}_{mixed}&=&\langle i, \{P^{+}_{Z}\},\{P^{-}_{Z}\},\{Z_p\} \rangle,\\
P^{\pm}_{Z}&\equiv& \frac{I\pm X_{p}\sigma^x_{p,p'}X_{p'}}{2},
\end{eqnarray}
where the gauge group $\mathcal{G}_{mixed}$ is non-Abelian.

The set of the stabilizers of the decohered state $\rho_{D}$ is given by taking the center of $\mathcal{G}_{mixed}$, $\mathcal{Z}(\mathcal{G}_{mixed})\equiv \mathcal{S}_D$. 
The decohered state is given by $\rho_D=\prod_{\ell}\frac{I-g^{D}_\ell}{2}$ where $g^D_{\ell}$ is a generator of $\mathcal{S}_D$.

Even for the decohered stabilizer set $\mathcal{S}_D$, logical operators can be found. 
The bare logical operators are the same as the original ones, $L_x$ and $L_y$. 
Thus, the original pure subsystem code (logical operators) is invariant. 
The center $\mathcal{S}_D$ is more elucidated from the viewpoint of the TC picture we shall show later.


\subsection{Mapped decoherence in TC system}
In the previous section, we showed that the LGHM can be mapped into the TC with rough and smooth boundaries. 
In what follows, we consider the effects of the gauge-operator decoherence from the TC description.

In the TC picture, the gauge-operator decoherence $\mathcal{E}^g$ becomes the following \cite{Nielsen2011}
\begin{eqnarray}
&&\mathcal{E}^{g}_{TC}=\prod_{\ell \notin \mbox{smooth} }\mathcal{E}^g_{TC,\ell},\\
&&\mathcal{E}^g_{TC,\ell}[\rho]=(1-p_x)\rho+p_x \sigma^x_{\ell}\rho \sigma^x_{\ell}.
\label{decoX_TC}
\end{eqnarray}
Note that this decoherence does not act on the link on the smooth boundaries.
From the relation of the Hilbert spaces, since the gauge operator decoherence $\mathcal{E}^{g}$ acts on the gauge-junk space, its local $\sigma^x$-decoherence in the TC system acts on the space $\bar{C}_{TC}^{(P,S_Z)}$. While the topological stabilizer code in the TC is robust in the decoherence $\mathcal{E}^{g}_{TC}$,  
the decoherence $\mathcal{E}^{g}_{TC}$ has a drastic impact on the gauge qubits. For periodic torus geometry, the effects of the decoherence $\mathcal{E}^{g}$ have been extensively investigated by employing the mapping to a statistical mechanical model \cite{JYLee2023,Zhu2023,Fan2024,Chen2024,wang2025_dec}.


\subsection{Decohered TC in zero-magnetic field limit}
We further discuss the state in the space $\bar{C}_{TC}$ orthogonal to the code space $C_{TC}$.
In this subsection, we consider the case of the zero magnetic field limit ($J=0$). The TC becomes an integrable stabilizer system corresponding to the zero-Higgs coupling limit in the LGHM.

First, we discuss $p_x=1/2$ case. The decoherence $\mathcal{E}^{g}_{TC}$ becomes non-selective projective measurement $\sigma^x_{\ell}$. Then, the gauging out approach is applicable. The gauge group is 
\begin{eqnarray}
\mathcal{G}^{TC}_{mixed}
&=&\langle i, \{P^{\pm}_{\sigma,\ell}\},\{ \tilde{B}_p\} \rangle,\\
P^{\pm}_{Z}&\equiv& \frac{I\pm \sigma^x_{\ell}}{2},
\end{eqnarray}
where the links $\ell$ on the smooth boundaries are not included. 
From this gauge group $\mathcal{G}^{TC}_{mixed}$, by following \cite{Ellison2023}, we can find the center as
\begin{eqnarray}
\mathcal{Z}(\mathcal{G}^{TC}_{mixed})&=&\langle i,\{ \tilde{A}_v\} \rangle.
\end{eqnarray}
Even for the stabilizer group $\mathcal{Z}(\mathcal{G}^{TC}_{mixed})$, the logical operator $L_{x(z)}$ remains while the $m$-anyon proliferates. 
The theory is only $e$-anyon, which is transparent. 
From these situations, the stabilized mixed state can be regarded as non-modular $Z^{(0)}_2$ topological order realized on the open-boundary system \cite{Ellison2023}.

We next investigate the behavior of the mixed state on an arbitrary $p_x$ line for the $J=0$ limit.
So far, the behavior of the TC set on a torus has been analyzed in the procedure proposed in \cite{JYLee2023}. 
We here apply it to the open boundary case. 
We start to pick up a single TC ground state (a logical state) as 
$\rho_{TC}=|\psi_{\alpha}\rangle\langle \psi_{\alpha}|$ where $|\psi_{\alpha}\rangle$ is one of the representative states of the single encoded logical 
qubit labeled by $\alpha$ in $\mathcal{L}_{TC}$. 
We consider to apply the decoherence $\mathcal{E}^g_{TC}$ to the state and obtain $p_x$-dependent decohered state as 
$\rho_{TC,D}=\mathcal{E}^{g}[\rho_{TC}]$. 
Here, we assume a sufficiently large system (with boundaries). 
By following the analytical treatment \cite{JYLee2023}, the decohered density matrix $\rho_{TC,D}$ is approximately given 
by the contribution of the partition function of the random-bond Ising model (RBIM) \cite{nishimori01,nishimori2011elements}
\begin{eqnarray}
&&\rho_{TC,D}= 
\sum_{m}\rho_{m}
\frac{\Lambda({\bf s}_m)}{2^{N+N_{BV}}(\cosh \beta)^{N_{BL}}}Z'_{RBIM}(\bf{s}_m,\beta_{\ell}),
\nonumber\\ &&\label{rho_D_stat}\\
&&Z'_{RBIM}(\bf{s},\beta)\nonumber\\
&&=\sum_{\{\sigma\}}\exp \biggr[-\beta\sum_{(v,v')\notin \mbox{on smooth}}s_{(v,v')}\sigma_v\sigma_{v'}\biggl],
\label{RBIM}
\end{eqnarray}
where 
$N$ is the total number of links, $N_{BV}$ is the total number of vertex sites except ones on smooth boundaries, 
and $N_{BL}$ is the total number of the link except ones on smooth boundaries. 
The factor $\Lambda({\bf s}_m)$ is a smooth boundary factor only taking a positive constant depending on the configuration $m$ 
the detailed form of which is shown in Appendix. 
The $p_x$-dependence is included in the inverse temperature $\beta$ by the relation $\tanh \beta=1-2p_x$ and $\rho_{m}$ is 
a block element of the density matrix constructed from basis states that belong to the same equivalent class denoted by $m$ \cite{JYLee2023}. 
$Z_{RBIM}(\bf{s},\beta)$ is the partition function of the RBIM of classical Ising spin $\{\sigma_{v}\}$ on the vertex $v$. 
${\bf s}$ is a link configuration and $s_{v,v'} \in {\bf s}$ represents a binary random coupling (taking $-1$ or $1$) 
defined the link between the vertex $v$ and $v'$. 
${\bf s}_m$ is a representative link configuration under an equivalence relation. 
In Appendix, we show the detailed calculations for Eqs.~(\ref{rho_D_stat}) and ~(\ref{RBIM}).

From Eqs.~(\ref{rho_D_stat}) and (\ref{RBIM}), we expect that the behavior of the decohered mixed state $\rho_{TC,D}$ 
is qualitatively governed by the behavior of the physics RBIM. 
The factor $\Lambda({\bf s})$ does not sweep the behavior of the physics RBIM.
Even for this boundary system, suppose the system is sufficiently large, the behavior of the mixed system is dominated by the bulk physics of the RBIM. 
In particular, the behavior of the critical phase transition of the RBIM can be inherited to the mixed state of the TC $\rho_{TC,D}$.

As a result, on $J=0$ line, 
the decohered matrix $\rho_{TC,D}$ reflects the behavior of the RBIM. 
Because the decoherence $\mathcal{E}^g_{TC}$ acts on the space $\bar{C}_{TC}^{(P,S_Z)}$, all changes originated from the RBIM 
are anticipated to occur the state in the space $\bar{C}_{TC}^{(P,S_Z)}$ in the TC.
In particular, the RBIM has a phase transition \cite{nishimori01,nishimori2011elements}. 
The critical point is denoted by $p_c$. 
The transition corresponds to the mixed state phase transition induced by the decoherence $\mathcal{E}^g_{TC}$ in the TC side, 
especially in the space $\bar{C}_{TC}^{(P,S_Z)}$. Here, by following Ref.~\cite{JYLee2023}, 
the entropy of $\rho_{TC,D}$ exhibits the clear phase transition at $p_c\sim 0.1094$ \cite{Doussal1988} and also 
if one focuses on the purity of $\rho_{TC,D}$, a phase transition behavior occurs at $p_c\sim 0.178$ \cite{KW_2DIsing,Onsager}.

\begin{figure}[t]
\begin{center} 
\vspace{0.5cm}
\includegraphics[width=8.5cm]{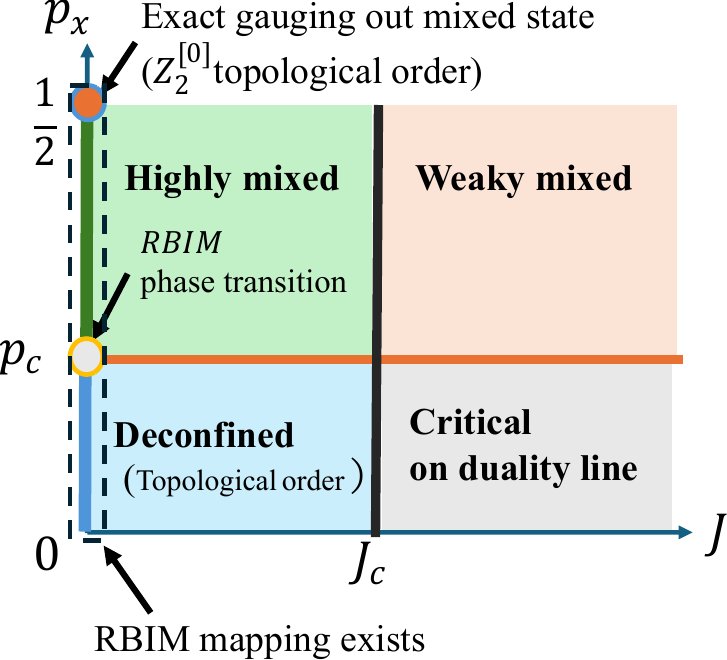}  
\end{center} 
\caption{Schematic phase structure in the bulk of the LGHM. 
The bulk gauge qubits in the gauge-junk space serve as the carrier of these pure and mixed phases. 
The name in the parentheses is on the side of the TC. 
The $J=0$ line can be understood by the behavior of the RBIM physics.}
\label{Fig3}
\end{figure}
Interestingly enough, this behavior of the mixed state depending on $p_x$ in the TC appears in the gauge qubits in the gauge-junk space 
$\bar{\mathcal{L}}^{(P,S_Z)}$ of the LGHM since the TC system is connected to the LGHM through the local unitary transformation $U_vU_p$. 
In particular, we expect that many critical behaviors, such as a divergence of the correlation length, large fluctuations for physical observables, 
and universality, etc., are inherited to the system of the LGHM. 
That is, the critical properties in the RBIM correspond to the presence of the critical mixed state of the gauge-junk space in the LGHM. 
\begin{figure*}[t]
\begin{center} 
\vspace{0.5cm}
\includegraphics[width=18cm]{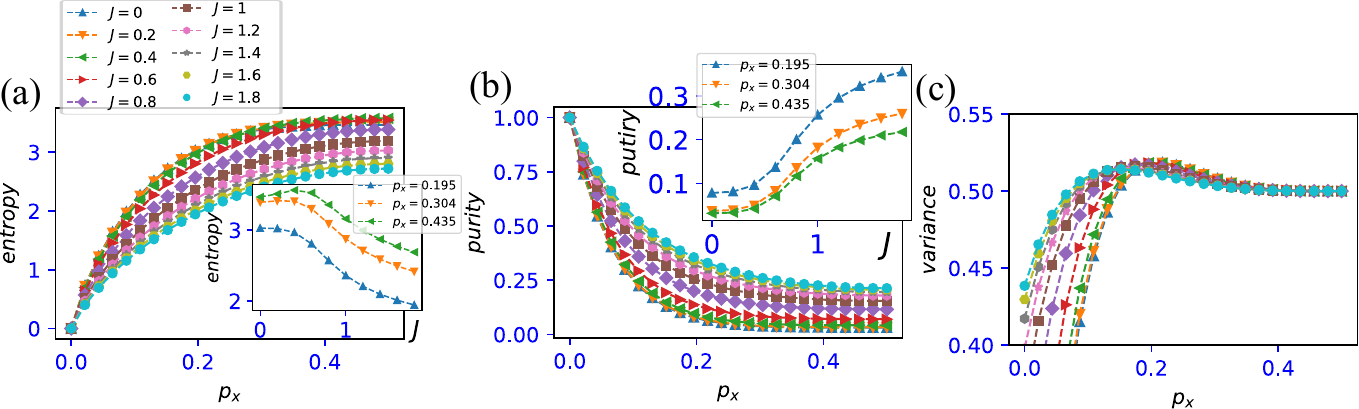}  
\end{center} 
\caption{(a) $p_x$- and $J$-dependence of the reduced entropy $S$. 
(b) $p_x$- and $J$-dependence for the reduced purity $P_r$. (c) Variance of the observable $\hat{O}^{TC}_{G}$. 
For each case, various cases of different $J$'s are investigated. 
$J$ determines the pure initial phase of the TC. 
The system size is $(L_x,L_y)=(3,2)$. 
The total number of spins is $N=13$.}
\label{Fig4}
\end{figure*}

Moreover, we note that the coexistence of a preserved logical space and a critical mixed state in such a subsystem code constitutes, 
in itself, a very interesting example of an exotic mixed state in open quantum systems.

\subsection{Global phase stricture of subsystem code}
Combining the previous result for the $J=0$ case, we draw the qualitative global phase diagram for this subsystem code in the LGHM 
under the decoherence $\mathcal{E}^g$ that is shown in Fig.~\ref{Fig3}. 
Each phase and transition line or point in this system is carried on by the gauge qubits in the gauge-junk space $\bar{\mathcal{L}}^{(P,S_Z)}$. 

So far, the $p_x=0$ line has been extensively studied; there are two phases, the deconfined phase (corresponding to topological order in the TC picture) 
and the critical phase due to the electro-magnetic duality  \cite{Trebst2008,PhysRevB.79.033109,PhysRevB.80.081104,Tupitsyn2010,Wu2012,Zhang2022_TC,Xu_2025}. 
We expect that there is a single transition point at $J=J_c$ due to the competition between the deconfined 
and critical phases \cite{Fradkin1979}, the deconfined phase is stable for $J< J_c$ and, for $J>J_c$, the critical phase appears.
We additionally comment that the first-order phase transition line can terminate at a certain value of $J>J_c$ in the GHM phase diagram, 
although it is not shown in the phase diagram in Fig.~\ref{Fig3}. 
For the sufficiently large $J$ regime, confinement and Higgs are the same phase \cite{Trebst2008,PhysRevB.79.033109,PhysRevB.80.081104,Tupitsyn2010,Xu_2025}.

Next, we turn to a finite $p_x$ case. 
We expect that the regime $p\lessapprox p_c$ and $J\lessapprox J_c$ (the right-blue regime in Fig.~\ref{Fig3}) can be the deconfined phase 
since the phase in $0\leq p_x \leq p_c$ on the $J=0$ line is connected to a regime with $p_x\lessapprox p_c$ and $J\lessapprox J_c$.
This connection can be understood by the scope of the Markov length, the inverse of which can be regarded as a gap of mixed 
state \cite{Sang2025,Zhang2024strong,negari2025spacetime,KOI_2025_ML}. 
If the Markov length does not diverge continuously between two mixed states, these mixed states belong to the same phase. 
Then, we expect that the phase transition point of the RBIM on the $J=0$ line is extended in a finite $J$ regime, that is, 
a phase transition line (the orange line) exists as shown in Fig.~\ref{Fig3}.
The transition line holds for large $J$, as is numerically verified later [See Fig.4(c)].

Then, we expect that the pure state phase transition point at $p_x=0$ and $J=J_c$ also extends in a finite $p_x$ region.
The critical transition line exists (the black line in Fig.~\ref{Fig3}). 
This elongates to a large $p_x$ [which is numerically verified later. [See Fig.4(a) and (b)].]
Again, under the assumption of the mixed phase equivalence through the Markov length, the regime for 
$p_x\lessapprox p_c$ and $J\gtrapprox 1$ (the right-gley regime in Fig.~\ref{Fig3}) can be the critical phase. 
Moreover, for large $p_x\gtrapprox p_c$ and small $J$ regime (the right-green regime in Fig.~\ref{Fig3}), 
we expect that a highly mixed state appears, which can be connected to the exact gauging out mixed state 
(corresponding to the $Z^{[0]}_2$ topological order \cite{Ellison2023} in the TC picture).
Finally, for large $p_x\gtrapprox p_c$ and large $J$ regime (the right-orange area in Fig.~\ref{Fig3}), 
we expect that another mixed phase exists, where a weakly mixed state with small entropy appears. 
As a result, the two extended transition lines separate four phases as shown in Fig.~\ref{Fig3}.

\section{Numerical study for small system}

In this section, we investigate the decoherence effect on the TC with the open boundary
by employing exact diagonalization by QuSpin package \cite{Quspin1, Quspin2} for the system with 
the total number of links $N=13$, i.e., $(L_x,L_y)=(3,2)$ of the vertex lattice.
We numerically observe entropy, $S=-\Tr[\rho_D\log \rho_D]$ and purity $P_r=\Tr[(\rho_D)^2]$.
As an efficient approximate calculation, we select the $300$ eigenvalues of the density matrix in descending order and 
obtain the reduced $S$ and $P_r$. 
Figures \ref{Fig4}(a) and \ref{Fig4}(b) are $p_x$ and $J$ behaviors for the reduced $S$ and $P_r$. 
The reduced entropy monotonically increases for various $J$. Since the finite-size effect is dominant due to the small size, 
the singular behavior is swept out. 
But we can observe the change in the system between the weak and strong decoherence. 
The system goes to a highly mixed phase for large $p_x$. 
The same behavior appears in $P_r$ shown in Fig.~\ref{Fig4}(b). 
It also exhibits a monotonic decrease for increasing $p_x$. 
For $p_x \gtrapprox 0.2$, the purity for various $J$ is almost saturated. 
This implies the state drastically changes for $p_x \gtrapprox 0.2$. 
From these data, in both opposite limits $p_x=0$ and $p_x=0.5$, states are drastically distinct. 
In both data sets, the strongest nonlinear behavior qualitatively appears around $p_x\sim 0.1$. 
This implies that this crossover is a finite-size precursor of the RBIM-type transition expected in the thermodynamic limit. 
We also shall show later that the variance of a gauge operator exhibits a remnant of a phase transition (the singular behavior of the data), 
which is a support for the remaining of the RBIM phase transition. 
We also show $J$-dependent behaviors of the reduced $S$ and $P_r$ in the inset of Fig.~\ref{Fig4} (a) and ~\ref{Fig4} (b). 
$S$ suddenly decreases around $J=0.8$ for any fixed $p_x$ as increasing $J$. This implies the existence of a phase transition 
(corresponding to the black solid line in Fig.~\ref{Fig3}). 
Also, $P_r$ suddenly increases around $J=0.8$ for any fixed $p_x$ as $J$ increases. 
This supports the existence of the phase transition.


\section{Stability for gauge qubit mixed subsystem code}

We discussed how gauge-junk space varies under the change in the model parameter and 
also the strength of the decoherence. 
For any phase of the gauge qubits, the subsystem code can be defined and the logical space exists. 

The strict separation between the logical space and the gauge-junk space is possible; 
however, the mixed phase of the gauge-junk space is \textit{not entirely irrelevant} to the stability of 
the logical qubit.
We here investigate the effect of the mixed phases of the gauge-junk space on the logical space. 
To this end, we observe the stability of the subsystem code against some kind of 
unitary perturbative coupling between the gauge qubits and the logical qubit. 
We find that states of the gauge qubit have a significant influence on the logical-qubit stability. 

To investigate the above issue, we perform a case study using 
a time evolution operator (for a small time interval), which describes interactions between 
the logical space and gauge-junk space, and is defined as, 
\begin{eqnarray}
U^{LG}_{\Delta t}=e^{-i\Delta t V_{LG}},\\
V_{LG}=\hat{L}_0\otimes \hat{O}_{G},
\end{eqnarray}
where $\Delta t$ represents a small time interval, $\Delta t\ll 1$, $\hat{L}_0$ is a specific logical operator 
and $\hat{O}_{G}=I_{L}\otimes (U_{G})$ acts only on the gauge-junk space. 
Here, we further impose the following requirements: 
(I) The operator $\hat{O}_{G}$ anti-commutates with the decoherence gauge operator (corresponding to $\sigma^x_\ell$). 
(II) The initial logical state of the subsystem code is \textit{not diagonal} with respect to the logical operator $\hat{L}_0$.

Then, we consider applying this short-time evolution to a decohered subsystem code $\rho_D$, 
and the interaction $V_{LG}$ starts to deform the encoded logical state. We observe 
how the entire state $\rho^{L}_D$ deforms depending on the mixed state of the gauge qubits in the gauge-junk space. 
Explicitly, we start to set the decohered state $\rho^{L}_D$ where the gauge qubits are in a mixed state, which is written by
\begin{eqnarray}
\rho^L_D=(|\psi_L\rangle \langle \psi_L|) \otimes \rho_{DG},
\end{eqnarray}
where we assume that the requirement of (II) is satisfied, that is, we choose the logical state not diagonal with respect to 
the operator $\hat{L}_0$ 
and $\rho_{DG}$ is a decohered mixed state of the junk qubits in the LGHM under the decoherence $\mathcal{E}^g$.

Here, we apply the time evolution operator $U^{LG}_{\Delta t}$ to $\rho^L_D$ and observe only the logical 
space by tracing out the gauge-junk space,
\begin{eqnarray}
\rho^{L'}_D=\Tr_{\rm junk}[U^{LG}_{\Delta t}\rho^L_D U^{LG,\dagger}_{\Delta t}].
\end{eqnarray}
$\rho^{L'}_D$ is further described as
\begin{eqnarray}
\rho^{L'}_D&=&P^{L_0}_{+}(|\psi_L\rangle \langle \psi_L|)P^{L_0}_{+}
+P^{L_0}_{-}(|\psi_L\rangle \langle \psi_L|)P^{L_0}_{-}\nonumber\\
&+&P^{L_0}_{+}(|\psi_L\rangle \langle \psi_L|)P^{L_0}_{-}\cdot\biggl[\Tr_{\rm junk}[e^{-i2\Delta t \hat{O}_G}\rho_{DG}]\biggr]\nonumber\\
&+&P^{L_0}_{-}(|\psi_L\rangle \langle \psi_L|)P^{L_0}_{+}\cdot\biggl[\Tr_{\rm junk}[e^{i2\Delta t \hat{O}_G}\rho_{DG}]\biggr],
\end{eqnarray}
where $P^L_{\pm}=\frac{I\pm \hat{L}_0}{2}$ and we used $\Tr_{\rm junk}[\rho_{DG}]=1$. 
Thus, we evaluate the short-time deviation of the logical encoded state in the logical space as
\begin{eqnarray}
&&\Delta\rho^{L}_D \equiv  \rho^{L'}_D-\Tr_{\rm junk}[\rho^L_D]\nonumber\\
&&=\begin{bmatrix}
0 & F(\Delta t,\hat{O}_G,\rho_{DG}) \\
F^*(\Delta t,\hat{O}_G,\rho_{DG}) & 0 \\
\end{bmatrix},\\
&&F(\Delta t,\hat{O}_G,\rho_{DG})
=\Tr_{\rm junk}[e^{-i2\Delta t \hat{O}_G}\rho_{DG}]-1.
\end{eqnarray}
The deformation appears in the off-diagonal term in terms of the operator $\hat{L}_0$. 
The off-diagonal element is further represented as 
\begin{eqnarray}
&&F(\Delta t,\hat{O}_G,\rho_{DG})\nonumber\\ 
&&\sim e^{-i2\Delta t\langle \hat{O}_G\rangle_J}[1-2(\Delta t)^2{\rm Var}[\hat{O}_G]]-1,
\end{eqnarray}
where
\begin{eqnarray}
\langle \hat{O}_G\rangle_J&\equiv&\Tr_{\rm junk}[\hat{O}_G \rho_{DG}],\nonumber\\ 
{\rm Var}[\hat{O}_G]&\equiv&\langle (\hat{O}_G)^2\rangle_J-\langle \hat{O}_G\rangle^2_J.
\end{eqnarray}
Here we used the cumulant expansion and took the term up to $\mathcal{O}(\Delta t^2)$. 
The off-diagonal in the deviation $\Delta \rho^{L}_D$ is related to the expectation value $\langle \hat{O}_G\rangle$ 
and the variance of the states calculated in the gauge-junk space. 
In particular, a large variance can induce a large deviation even for a short-time evolution. 
This off-diagonal deviation implies that the state of the gauge qubit in the gauge-junk space gives a significant impact on the stability of 
the encoded qubit of the subsystem code. If the coupling represented by the time evolution $U^{LG}_{\Delta t}$ is a perturbation 
from the environment, 
the encoded qubit gradually changes regardless of the type of gauge qubit. 
However, in the case where the variance becomes significantly large, the change becomes more pronounced. 
This suggests that when the gauge qubit state is in a critical mixed state where the variance for a gauge operator $\hat{O}_G$ is large, 
the logical qubit suddenly loses its coherence, that is, its stability is weak.

We also comment on the choice of $\hat{O}_G$. 
The variance depends on the model parameters in the LGHM and its pure ground state phases. Generally, there are some candidates as $\hat{O}_G$ leading to a large variance. 
As an example, we consider 
\begin{eqnarray}
\hat{O}_G = \frac{1}{N_p}\sum_{p\in \mbox{some set}} B_{p},
\end{eqnarray}
which is anticommute with the decoherence operator $\sigma^x_{\ell}$, and the range of the sum works as a parameter that controls 
the magnitude of the variance. 

Under this setup, we observe the absolute value of the off-diagonal element $|F(\Delta t,\hat{O}_G,\rho_{DG})|$. 
If $\rho_{DG}$ is a critical mixed state, the variance ${\rm Var}[\hat{O}_G]$ is large since a general critical mixed state is 
expected to have a large value of ${\rm Var}[\hat{O}_G]$. 
The fluctuation of the absolute off-diagonal part is getting large in the time evolution.
On the other hand, in the case that $\rho_{DG}$ is a stable 
mixed order, the value of ${\rm Var}[\hat{O}_G]$ is small compared to one in the critical state. The fluctuation of the absolute 
off-diagonal part is small.

Overall, the weak interaction considered here, which spans both the logical space and the gauge qubits, acts in a direction 
that disrupts the encoded qubit of the subsystem code. 
In particular, when the gauge-junk space is in a critical mixed state, the effects of the interaction become pronounced. 
Although the interaction we considered is a rather specific one, it may also be regarded as a representative case of 
an external perturbation coming from the environment.

Finally, as a verification, we numerically observe the variance in the small-size system, the same as the previous one by using exact 
diagonalization. 
As an example, 
we pick up the following gauge operator interaction
\begin{eqnarray}
\hat{O}_G = \frac{1}{2}(Z_{p_1}+Z_{p_2}).
\end{eqnarray}
Here for the interaction, we turn to the TC representation such as 
\begin{eqnarray}
\hat{O}_G = \frac{1}{2}(Z_{p_1}+Z_{p_2}) 
\xrightarrow[\text{gauge fixing}]{U_{v}U_{p}}\hat{O}^{TC}_G = \frac{1}{2}(B_{p_1}+B_{p_2}).\nonumber\\
\end{eqnarray}
We calculate ${\rm Var}[\hat{O}^{TC}_G]$ for various model parameters in the TC picture, the same as the previous numerics. 
Note that ${\rm Var}[\hat{O}^{TC}_G]$ in the TC coincides with ${\rm Var}[\hat{O}_G]$ in the LGHM.

The behavior is shown in Fig.~\ref{Fig4}(c). Even in this small system, the variance exhibits a soft peak around $p_{x}=0.15$ for various $J$. This is a signature of the mixed state phase transition in the bulk. Around the soft peaks, the variance is large. 
Thus, the large value induces the large fluctuation of the off-diagonal element in $\Delta \rho^L_D$. 
This numerical result supports our prediction for the stability of the encoded qubit to the interaction between the gauge qubit and logical space. 

In addition, as a by-product of this calculation, this behavior of the variance supports the prediction of the presence of the mixed state phase transition related to the RBIM.

\section{CONCLUSIONS}
In this work, the subsystem code of the generic LGHM proposed in the recent works \cite{Wildeboer2022} was investigated. 
We especially studied the effect of a type of decoherence on the LGHM, where the operator of the decoherence is a gauge operator. 
The decoherence is specific in a sense that it affects only the gauge qubits, but not the logical sector. 
The combination of the decoherence strength and the set of the model parameters leads to the rich mixed phases in the gauge qubits. 
We clarified the presence of various subsystem codes with different gauge-qubit mixed states by mapping the model to the TC code 
under the decoherence, which has been extensively investigated in \cite{JYLee2023}. 
We further drew out the global mixed-phase diagram of the subsystem code. 
In the large decoherence limit, the subsystem code can be understood as a mixed subsystem code constructed by the gauging out 
procedure\cite{Ellison2023,Sohal2025}. 
Indeed, the subsystem code in the LGHM under the measurements of gauge operators is a concrete example of the mixed subsystem code 
recently proposed in \cite{Sohal2025}. 
Various states with the encoded qubit in the LGHM are interesting from the viewpoint of the quest for unconventional mixed states 
in decohered many-body systems. 
Especially, the subsystem code with mixed critical bulk that we recognized is just a co-existence of logical operator 
(a non-local boundary symmetry) and bulk mixed critical, where we expect the vanishing of the Markov gap \cite{Sang2025,Zhang2024strong}. 
We also discussed the stability of the subsystem code in the LGHM. 
The subsystem code, of course, is well-defined for any phases in the LGHM under the gauge-operator decoherence, 
independent of the kind of gauge-qubit mixed state. 
However, from the viewpoint of the stability of the subsystem code, the mixed state of the gauge qubits is important. 
In this work, by introducing a specific dynamical perturbation, we have evaluated how the stability of subsystem codes is 
affected when they coexist with different mixed-state phases. 
Our analysis reveals that the emergence of a critical mixed state of gauge qubits in the LGHM degrades the stability of the subsystem code.
Overall, these findings indicate that subsystem codes embedded in critical or mixed-state environments are generally more 
susceptible to environmental perturbations than those associated with gapped mixed-state phases.

As future interesting topics, we can point out the following: 
(i) How the subsystem code considered here is related to the decoherence-free quantum code \cite{Lidar1998,Lidar2003} in a more strict and mathematical 
description. 
A detailed investigation of how the decohered subsystem code we have observed here relates to the previously established framework of 
decoherence-free codes\cite{Lidar1998,knill2000,Lidar2003,Kribs2005}. 
(ii) In this work, we focus on the ground state physics as the starting point. 
However, the Hilbert space structure of the subsystem code in this model keeps even for highly excited states\cite{Wildeboer2022,KI2023_sc}. 
Essentially, this can be well-understood by the notion of strong zero mode~\cite{Fendley_2016,Else_2017}. 
How the notion harmonizes with the subsystem code under the decoherence studied in this work is a future-interesting issue.

\section*{Acknowledgements}
This work is supported by JSPS KAKEN-HI Grant Number 23K13026 (Y.K.). 

\renewcommand{\thesection}{A\arabic{section}} 
\renewcommand{\theequation}{A\arabic{equation}}
\renewcommand{\thefigure}{A\arabic{figure}}
\setcounter{equation}{0}
\setcounter{figure}{0}
\appendix
\widetext
\section*{Appendix: Mapping to a statistical physics model}
The density matrix of the TC with rough and smooth boundaries under the decoherence $\mathcal{E}^g_{TC}$ can be mapped into the partition function of the RBIM 
for a sufficiently large system size. 
By following Ref.~\cite{JYLee2023}, we here show the mapping in detail. 
We start with the case of the infinite size of the system and then move to the case of the open boundary case.

\subsection*{System of an infinite size: without boundary case}

To treat the state of the TC under a decoherence $\mathcal{E}^g_{TC}$, 
we introduce a set of basis given as follows,
\begin{eqnarray}
|\Omega_{\bf s}\rangle =\prod_{\ell}(\sigma^{x}_{\ell})^{s_\ell}|\uparrow\rangle^{\otimes N},
\end{eqnarray}
where $s_{\ell}$ is a classical bit variable (taking $-1$ or $1$) and $|\uparrow\rangle$ is $z$-component spin-$1/2$ basis, $Z|\uparrow(\downarrow)\rangle=1(-1)|\uparrow(\downarrow)\rangle$. 
$N$ is the total number of the links of the TC system. 
${\bf s}$ represents a configuration of each $s_{\ell}$, also referring the label of the basis of the system.

Here, we consider a pure ground state of the TC (in an infinite system). 
Due to no boundary or no specific type of geometry, the ground state is unique \cite{Pachos2012,Wen_text}. 
The density matrix is given by
\begin{eqnarray}
\rho_{TC}= \prod_{v,p}\biggl[\frac{1+\tilde{G}_v}{2}\biggr]\biggl[\frac{1+\tilde{B}_p}{2}\biggr] \equiv|\psi_{TC}\rangle\langle \psi_{TC}|.
\end{eqnarray}
We next apply the decoherence $\mathcal{E}^g_{TC}$ to $\rho_{TC}$, where the decoherence acts on all links. 
Then, we can write down the matrix element of the decohered density matrix in terms of the basis $|\Omega_{\bf s}\rangle$ as \cite{JYLee2023} 
\begin{eqnarray}
&&\langle\Omega_{\bf s}|\mathcal{E}^{g}_{TC}[\rho_{TC}]|\Omega_{\bf s'}\rangle=\langle \psi_{TC}|\mathcal{E}^g_{TC}[|\Omega_{\bf s'}\rangle\langle\Omega_{\bf s}|]|\psi_{TC}\rangle=\frac{1}{2^N}\sum_{\ell_c}(1-2p)^{|\ell_c|}\biggr[\prod_{\ell\in\ell_c}s_{\ell}\biggl]K(\ell_c,{\bf s},{\bf s'}),
\end{eqnarray}
where $\ell_c$ represents the label of a link configuration for $s_{\ell}$ and $|\ell|_c$ is the number of the links in the configuration $\ell_c$ and 
\begin{eqnarray}
K(\ell_c,{\bf s},{\bf s'})&=&\langle \psi_{TC}|\prod_{\ell\in \ell_c}\sigma^z_{\ell}\prod_{\ell:all}(\sigma^{x}_{\ell})^{(1-s_{\ell}s'_{\ell})/2}\|\psi_{TC}\rangle
=\delta_{\partial{\ell_c},0}\times \delta_{\partial ({\bf s}\cdot{\bf s'}),0}.
\end{eqnarray}
Here, the factor $\delta_{\partial{\ell_c},0}$ means that if $\ell_c$ has no boundary ($\partial \ell_c=0$), it becomes $1$, otherwise zero and 
the factor $\delta_{\partial ({\bf s}\cdot{\bf s'}),0}$ has the same meaning, where
${\bf s}\cdot{\bf s'}$ denotes the element-wise product of the two variable configurations.
The above condition means that the ground state $|\psi_{TC}\rangle$ has no open $\sigma^x$ or $\sigma^z$ string \cite{Wen_text}, 
if one applies an open $\sigma^x$ or $\sigma^z$ string to $|\psi_{TC}\rangle$, 
then $e$ and $m$ excitations create the state of which is an orthogonal excited state to $|\psi_{TC}\rangle$.

The matrix element above is further transformed as 
\begin{eqnarray}
\langle\Omega_{\bf s}|\mathcal{E}^{g}_{TC}[\rho_{TC}]|\Omega_{\bf s'}\rangle
&=&\frac{\delta_{\partial({\bf s}\cdot{\bf s'}),0}}{2^N}\sum_{\ell_c\in (\mbox{loop config.})}(1-2p)^{|\ell_c|}\biggr[\prod_{\ell\in\ell_c}s_{\ell}\biggl]\nonumber\\
&=&\frac{\delta_{\partial({\bf s}\cdot{\bf s'}),0}}{2^{3N/2}(\cosh \beta)^{N}}2^{N/2}(\cosh \beta)^{N}\sum_{\ell_c\in (\mbox{loop config.})}(\tanh \beta)^{|\ell_c|}\biggr[\prod_{\ell\in\ell_c}s_{\ell}\biggl]\nonumber\\
&=&\frac{\delta_{\partial({\bf s}\cdot{\bf s'}),0}}{2^{N/2}(2\cosh \beta)^{N}}Z_{RBIM}(\bf{s},\beta),
\end{eqnarray}
where 
\begin{eqnarray}
Z_{RBIM}(\bf{s},\beta)&=&\sum_{\{\sigma\}}\exp \biggr[-\beta\sum_{(v,v')}s_{(v,v')}\sigma_v\sigma_{v'}\biggl]
\end{eqnarray}
with $\tanh \beta=1-2p_x$. In the above equation, $\sigma_{v}$ represents a classical Ising spin on the vertex $v$ and $s_{v,v'}$ 
represents a binary random coupling defined on the link between the vertex $v$ and $v'$. 

Consequently, the decohered density matrix is related to the partition function of the RBIM. 
The behavior of the RBIM depending on the inverse temperature $\beta$ is reflected in that of the decohered density matrix of the TC system.

\subsection*{Open boundary case}
Based on the without boundary case above, we turn to consider the open boundary case, mainly focused on in the main text.
We especially need to observe what the form of $\langle\Omega_{\bf s}|\mathcal{E}^{g}_{TC}[\rho_{TC}]|\Omega_{\bf s'}\rangle$ is. 
Along with the setup of the decoherence considered in the main text, the decoherence $\mathcal{E}^{g}_{TC}$ is required not to be applied on the links on the smooth boundaries. 
This point is treated as introducing the link-dependent $p$ as $p\to p_{\ell}$ and assuming the decoherence is applied to all links of the TC system with the boundaries. 
Finally, we tune the value of $p_{\ell}$ as $p_{\ell}=p_x (\ell \notin \mbox{smooth})$, $p_{\ell}=0 (\ell \in \mbox{smooth})$.

In this case, we focus on the pure logical $L_z=+1$ state denoted by $|\psi^{+z}_{TC}\rangle$, satisfying $L_z|\psi^{+z}_{TC}\rangle=(+1)|\psi^{+z}_{TC}\rangle$. 
We assume that the state is in $(\tilde{P},\tilde{S}_Z)=(+1,+1)$ sector. 
We denotes its density matrix by $\rho^o_{TC}\equiv |\psi^{+z}_{TC}\rangle \langle \psi^{+z}_{TC}|$. 
We again consider the matrix element of the decohered state from the initial state $\rho^o_{TC}$ is
\begin{eqnarray}
\langle\Omega_{\bf s}|\mathcal{E}^{g}_{TC}[\rho^o_{TC}]|\Omega_{\bf s'}\rangle&=&
\frac{1}{2^N}\sum_{ \ell_c}\biggr[\prod_{\ell\in \ell_c}(1-2p_{\ell})s_{\ell}\biggl]\times \langle \psi^{+z}_{TC}|\prod_{\ell\in \ell_c}\sigma^z_{\ell}\prod_{\ell:all}(\sigma^x_{\ell})^{(1-s_{\ell}\cdot s'_{\ell})/2}|\psi^{+z}_{TC}\rangle \nonumber\\
&\equiv& 
\frac{1}{2^N}\sum_{\ell_c}\biggr[\prod_{\ell\in \ell_c}(1-2p_{\ell})s_{\ell}\biggl]\times K^{o}(\ell_c,{\bf s},{\bf s}'). 
\end{eqnarray}
Here, $\ell_c$ is a link configuration. 
For the above equation, we firstly need to treat the operator product $\prod_{\ell:all}(\sigma^x_{\ell})^{(1-s_{\ell}\cdot s'_{\ell})/2}$ in $K^{o}(\ell_c,{\bf s},{\bf s}')$. 
Since the state $|\psi^{+z}_{TC}\rangle$ is the logical $L_z=+1$ state, it is sufficient to pick up $\partial({\bf s}\cdot{\bf s}')=0$ case (${\bf s}\cdot{\bf s}'$ is no boundary
configuration). 
Then, if ${\bf s}\cdot{\bf s}'$ is a loop configuration plus two operator strings picked up from $L_x$ and $L_x\tilde{P}$, such a configuration ${\bf s}\cdot{\bf s}'$ is 
the same contribution to the case of only loop configuration ${\bf s}\cdot{\bf s}'$. 
Thus, the scale of the partition function only changes. It gives no physical change.
From this point, the factor $\delta (\partial_{bulk}({\bf s}\cdot{\bf s}')=0)$ comes from $K^{o}(\ell_c,{\bf s},{\bf s}')$, where $\partial_{bulk}$ means to take 
the boundary for the configuration ${\bf s}\cdot{\bf s}'$, where if the configuration includes the support of the operator $\tilde{P}$ on the rough boundaries, 
the part of the boundary in the configuration is ignored. 
The factor $\delta (\partial_{bulk}({\bf s}\cdot{\bf s}')=0)$ also means that if $\partial_{bulk}({\bf s}\cdot{\bf s}')=0$, one takes one, otherwise zero.

We next focus on the operator product $\prod_{\ell\in \ell_c}\sigma^z_{\ell}$ in the factor $K^{o}(\ell_c,{\bf s},{\bf s}')$. 
We here identify what type link configurations $\ell_c$ make the factor $K^{o}(\ell_c,{\bf s},{\bf s}')$ non-zero. 
They are the following four types of the link configuration: 
\begin{itemize}
\item 
$\ell_c$ includes loops only. Its contribution is denoted by $K_0$.

\item 
$\ell_c$ includes loops plus one $L_z$ string configuration (on left smooth boundary). Its contribution is denoted by $K_1$.

\item 
$\ell_c$ is loops plus $L_z\tilde{S}_Z$ string configuration (on right smooth boundary). Its contribution is denoted by $K_2$.

\item 
$\ell_c$ is loops plus $\tilde{S}_Z$ string configuration (on right and left smooth boundaries). Its contribution is denoted by $K_3$.\\
\end{itemize}
Summering these contributions and putting forward the constraint factor for the configuration $({\bf s}\cdot {\bf s}')$, $\langle\Omega_{\bf s}|\mathcal{E}^{g}_{TC}[\rho^o_{TC}]|\Omega_{\bf s'}\rangle$ can be written as 
\begin{eqnarray}
\langle\Omega_{\bf s}|\mathcal{E}^{g}_{TC}[\rho^o_{TC}]|\Omega_{\bf s'}\rangle&\equiv&\delta (\partial_{bulk}({\bf s}\cdot{\bf s}')=0)\frac{1}{2^{N}}[K_0({\bf s})+K_1({\bf s})+K_2({\bf s})+K_3({\bf s})],
\end{eqnarray}
where 
\begin{eqnarray}
K_0({\bf s})&=&\sum_{\ell_c\in (\mbox{loop config.})}\biggr[\prod_{\ell\in\ell_c}(1-2p_{\ell})s_{\ell}\biggl]=\frac{1}{2^{N_{BV}}(\cosh \beta)^{N_{BL}}}Z'_{RBIM}(\bf{s},\beta_{\ell}),\\
K_1({\bf s})&=&\sum_{\ell_c\in (\mbox{loop config.+$L_z$ string})}\biggr[\prod_{\ell\in\ell_c}(1-2p_{\ell})s_{\ell}\biggl]=
\biggl(\prod_{\ell \in \mbox{left smooth}}s_{\ell}\biggr) K_0({\bf s}),\\
K_2({\bf s})&=&\sum_{\ell_c\in (\mbox{loop config.+$L_z\tilde{S}_Z$ string})}\biggr[\prod_{\ell\in\ell_c}(1-2p_{\ell})s_{\ell}\biggl]
=\biggl(\prod_{\ell \in \mbox{right smooth}}s_{\ell}\biggr) K_0({\bf s}),\\
K_3({\bf s})&=&\sum_{\ell_c\in (\mbox{loop config.+$\tilde{S}_Z$ string})}\biggr[\prod_{\ell\in\ell_c}(1-2p_{\ell})s_{\ell}\biggl]=\biggl(\prod_{\ell \in \mbox{left-right smooth}}s_{\ell}\biggr) K_0({\bf s})
\end{eqnarray}
and 
\begin{eqnarray}
Z'_{RBIM}(\bf{s},\beta_{\ell})&\equiv&\sum_{\{\sigma\}}\exp \biggr[-\beta\sum_{(v,v')\notin \mbox{on smooth}}s_{(v,v')}\sigma_v\sigma_{v'}\biggl].
\end{eqnarray}
Here, $N_{BV}$ is the total number of the vertex site except ones on smooth boundaries, $N_{BL}$ is the total number of the link except ones on smooth boundaries.
As a result, 
\begin{eqnarray}
&&\langle\Omega_{\bf s}|\mathcal{E}^{g}_{TC}[\rho^o_{TC}]|\Omega_{\bf s'}\rangle=\delta (\partial_{bulk}({\bf s}\cdot{\bf s}')=0)\frac{1}{2^{N}}\Lambda({\bf s})K_0({\bf s}),\\
&&\Lambda({\bf s}) \equiv 1+\biggl(\prod_{\ell \in \mbox{left smooth}}s_{\ell}\biggr)+\biggl(\prod_{\ell \in \mbox{right smooth}}s_{\ell}\biggr)+\biggl(\prod_{\ell \in \mbox{left-right smooth}}s_{\ell}\biggr).
\end{eqnarray}
We reach to the final form of the decohered density matrix, 
\begin{eqnarray}
\mathcal{E}^{g}_{TC}[\rho^o_{TC}]&=&\sum_{{\bf s},{\bf s}'}|\Omega_{\bf s'}\rangle \langle\Omega_{\bf s}|
\biggl[\delta (\partial_{bulk}({\bf s}\cdot{\bf s}')=0)\frac{1}{2^{N}}\Lambda({\bf s})K_0({\bf s})\biggr]
=\sum_{m}\rho_{m}
\frac{1}{2^{N}}\Lambda({\bf s}_m)K_0({\bf s}_m),
\end{eqnarray}
where $\rho_{m}=\sum_{{\bf s},{\bf s}':({\bf s} \sim {\bf s}')}|\Omega_{{\bf s}'}\rangle \langle \Omega_{{\bf s}}|$. 
The equivalence relation ${\bf s} \sim {\bf s}'$ comes from the condition $\partial_{bulk}({\bf s}\cdot{\bf s}')=0$ \cite{JYLee2023}. $\bf{m}$ is 
a representative under the equivalence class and its representative link configuration is ${\bf s}_{m}$. 
From this representation, the behavior of the elements $\langle\Omega_{\bf s}|\mathcal{E}^{g}_{TC}[\rho^o_{TC}]|\Omega_{\bf s'}\rangle$ is governed by $K_0$, 
that is, the behavior of the RBIM. 
The factor $\Lambda({\bf s})$ comes from no decoherence on the smooth boundaries. 
This, however, does not sweep the essential behavior of the RBIM physics. 
Also, we expect that the behavior is carried on the bulk part of the system.

From the form of the decohered state, we expect that the entropy and purity are affected by the behavior of the RBIM physics\cite{JYLee2023}.

\endwidetext
\clearpage
\bibliography{ref}

\end{document}